\newcommand{\gev}{\!\mathrm{GeV}}
\newcommand{\mev}{\!\mathrm{MeV}}
\newcommand{\kev}{\!\mathrm{keV}}
\newcommand{\cm}{\!\mathrm{cm}}

\newcommand{\nm}{\!\mathrm{nm}}

\documentclass[5p]{elsarticle}
\preprintfalse

\usepackage{amsmath}

\begin{document}
\begin{frontmatter}
\title{ Radiation Damage of F8 Lead Glass with $20~\mev$ Electrons }

\author[iu]{B.~D.~Schaefer}
\author[iu]{R.~E.~Mitchell}
\author[iuceem]{P.~McChesney}
\author[iu]{M.~R.~Shepherd\corref{cor1}}
\ead{mashephe@indiana.edu}
\cortext[cor1]{Corresponding author}
\author[iu]{J.~M.~Frye}
\address[iu]{Indiana University, Bloomington, Indiana 47405, USA}
\address[iuceem]{Center for the Exploration of Energy and Matter, Indiana University, Bloomington, IN 47408, USA}
\date{\today}

\begin{abstract}
Using a $20~\mev$ linear accelerator, we investigate the effects of electromagnetic radiation on the optical transparency of F8 lead glass.  Specifically, we measure the change in attenuation length as a function of radiation dose.  Comparing our results to similar work that utilized a proton beam, we conclude that F8 lead glass is more susceptible to proton damage than electron damage.
\end{abstract}


\end{frontmatter}

\section{Introduction}
Lead glass is a common absorber used in the construction of electromagnetic calorimeters for particle physics detectors.  One such device is the GlueX detector, which is under construction at Jefferson Lab.  The electromagnetic beam for the GlueX experiment will be derived from $12~\gev$ electrons accelerated by the Continuous Electron Beam Accelerator Facility (CEBAF) to produce $9~\gev$ linearly polarized photons via coherent bremsstrahlung radiation from a thin diamond wafer.  These photons will strike a proton target and the resulting reaction will be analyzed with the goal of searching for exotic hybrid mesons.  Production of such mesons with polarized photons adds an additional angular observable to enhance the capability of the experiment to determine the quantum numbers of the meson.  The search for these putative exotic mesons is one of the key goals of the GlueX experiment.

The GlueX forward calorimeter, designed to measure the energy of photons near the beam axis, will be made of 2800 F8 lead glass\footnote[1]{See Refs.~\cite{e852,radphi,selex} for more information on F8 lead glass and its use in previous detectors.} bars optically coupled to FEU 84-3 photomultiplier tubes.  It is well known that lead glass loses transparency when irradiated, and there is a need to understand the potential long term damage to the GlueX calorimeter from radiation associated with the beam.  Since the primary radiation source in the GlueX environment is electromagnetic, we irradiated F8 lead glass blocks with $20~\mev$ electrons produced by a linear accelerator at the Indiana University Center for the Exploration of Energy and Matter (IUCEEM).  The transmission through the damaged blocks was measured using a spectrophotometer.

In what follows, we first discuss our method of calibrating the electron flux of the accelerator that was used to irradiate the blocks.  Next, we discuss the development of a model for the transmission of light through the glass based on our analysis of the irradiated blocks.  We conclude this paper by comparing our results to those obtained in a different study that utilized a hadron beam.

 \section{Calibrating the Linear Accelerator}

\begin{figure}
\begin{center}
\includegraphics[width=\linewidth]{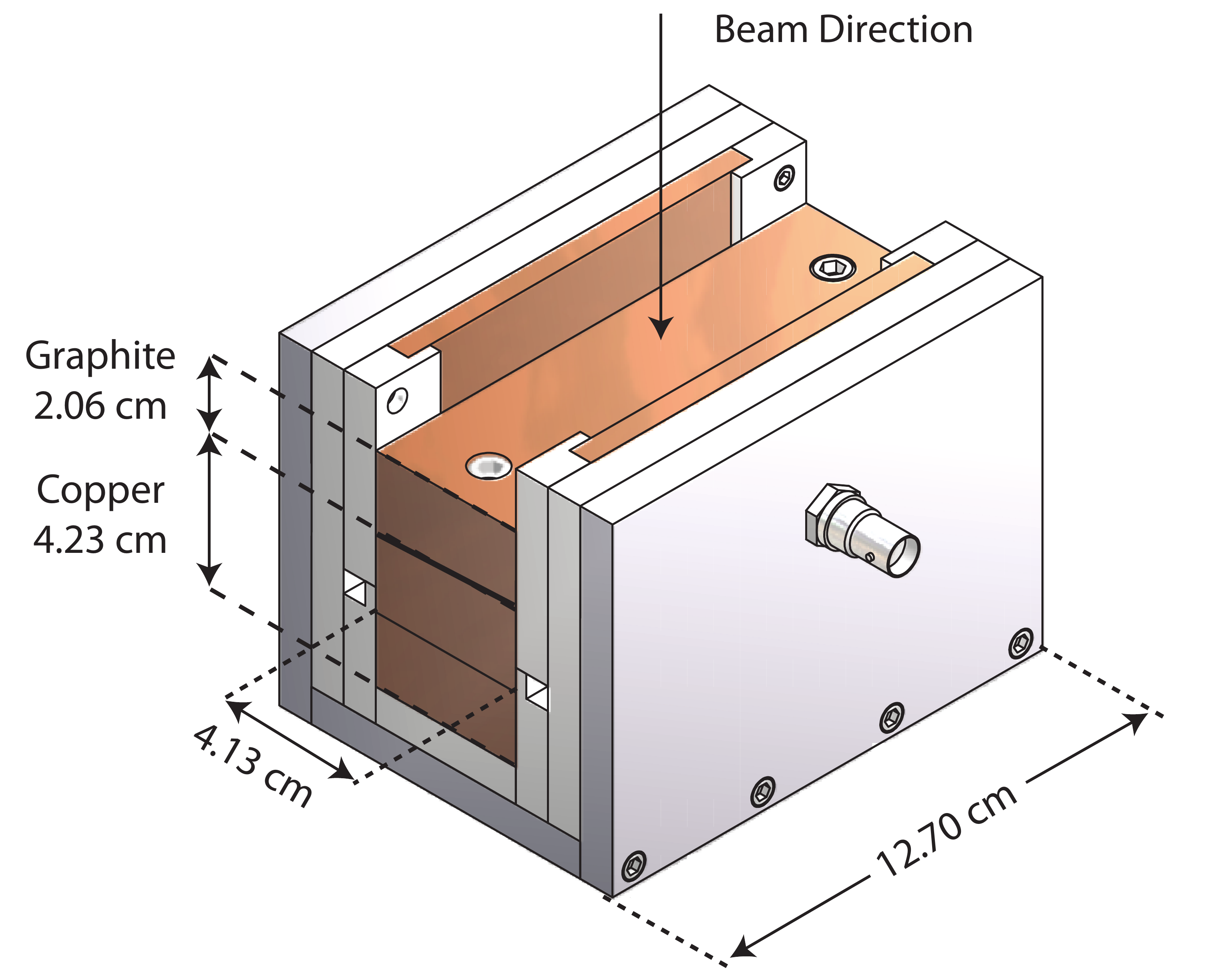}
\caption{Drawing of the Faraday cup used to calibrate the linear accelerator used in this study. }
\label{fig:FC_draw}
\end{center}
\end{figure}

The $20~\mev$ Varian Clinac electron accelerator delivers pulses that have a duration on the order of a few microseconds at a rate of about 33~Hz.  In order to determine the total energy deposited in the lead glass block it was necessary to first determine the total charge in a single pulse.  We then used a counting circuit to count the number of pulses delivered to the block, which ranged from about 4~\!900 to 88~\!000.

A Faraday cup was constructed (see Fig.~\ref{fig:FC_draw}) to measure the charge of a single pulse emitted by the accelerator.  A Faraday cup achieves this by having a core that is struck by the electron beam.  The core is grounded through a resistor, and the current leaving the cup can be measured to obtain the number of electrons that struck core.  The design of the Faraday cup was such that the core could be interchanged with a lead glass block; this allowed irradiation of the lead glass blocks in the same location as the measurement.  The Faraday cup was built with bias plates that could be set to a large negative voltage, with respect to the core, to reduce electron leakage from the core.

\begin{figure}
\includegraphics[width = \linewidth]{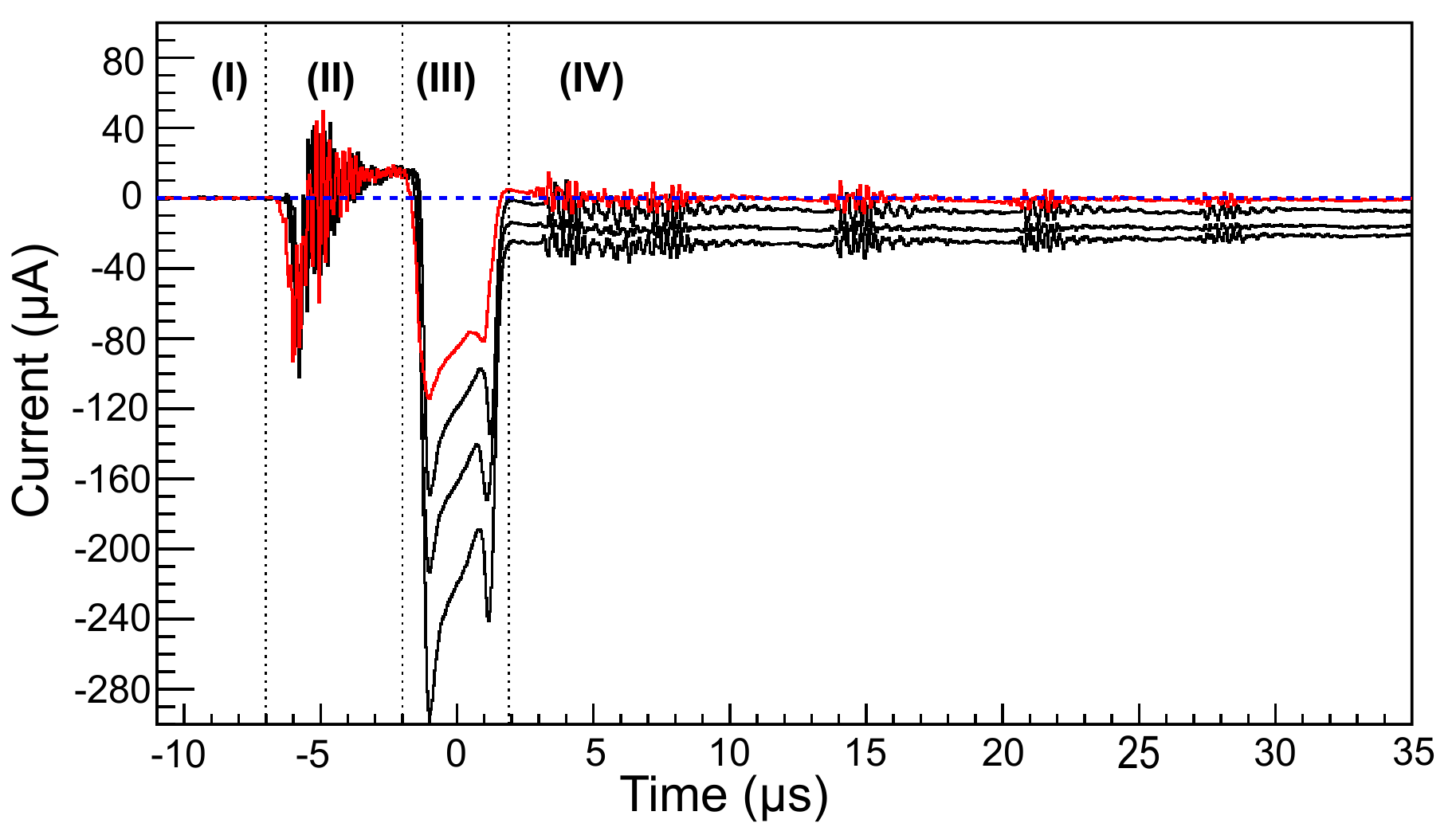}
\caption{Faraday cup output for different settings of the bias voltage.  The top (red) line is with no bias voltage.  The lines below this are -100~V, -300~V, and -500~V, respectively.  See text for a discussion of the Regions labeled I to IV.  }
\label{fig:6pulses}
\end{figure}

The Faraday cup current as a function of time is shown in Figure~\ref{fig:6pulses}.  The top solid line shows the results for zero bias voltage and each line below that uses a bias of -100~V, -300~V, and -500~V, respectively.  There are four regions of interest in the figure.  In Region~I the current is constant with a value that is independent of bias voltage.  Region~II contains RF noise from the accelerator.   Region~III is the actual pulse.  Region IV is a tail that is the result of electrons produced from ionization that are collected by the cup, aided by the bias voltage.  The structures in Region~IV are attributed to accelerator noise.

It has been previously noted that a Faraday cup can give erroneous measurements of the current~\cite{old_paper}.  The causes described are: (1) penetration of the electron shower out of the bottom and sides of the cup, (2) backscatter of the electrons which escape the core of the cup, (3) leakage of the current to ground, and (4) ions produced in the vicinity of the cup.  The penetration of the electron shower was reasonably accounted for in the design of the core.  The beam profile area was $2.4~\cm\times2.5~\cm$.  Electrons lose on average about $13~\mev$ per cm traveled in copper and $4~\mev$ per cm in graphite, so the dimensions of the core should capture most of the electrons.  The error due to backscatter was minimized by the inclusion of bias plates and a graphite core in the Faraday cup.  The graphite is used because it has a lower atomic number than copper and therefore is less susceptible to backscatter.  \textsc{Geant4}~\cite{geant} simulations show that for a bias voltage of zero, around 90\% of electrons above $1~\kev$ are captured by the cup.  Finally, Delrin is used as an insulator to minimize current leaking to ground.  

\begin{figure}
\includegraphics[width = \linewidth]{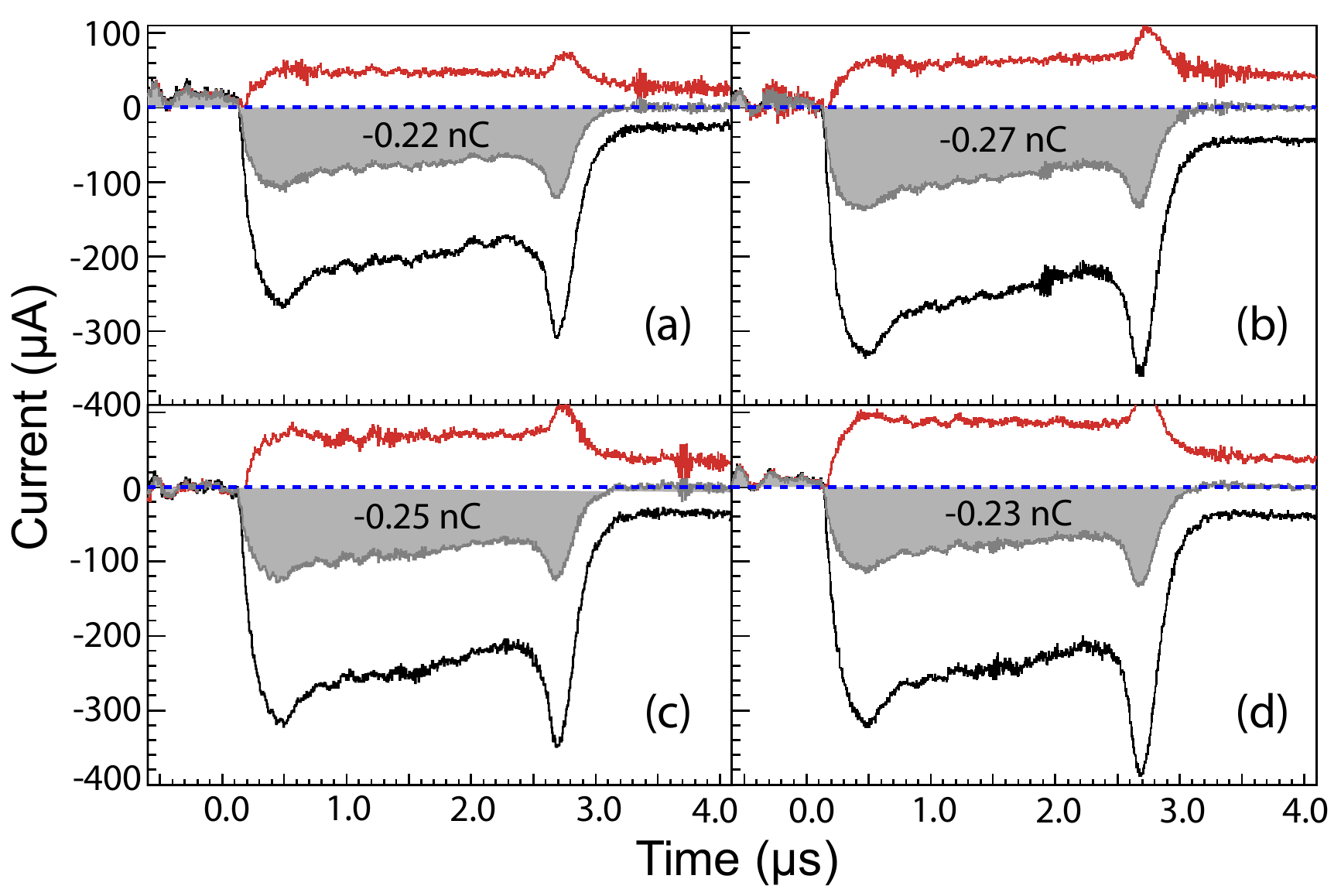}
\caption{Faraday cup output.  Each figure is taken with a different bias voltage.  The bias voltages are (a) $\pm600~\text{V}$,  (b) $\pm700~\text{V}$, (c) $\pm800~\text{V}$, and (d) $\pm900~\text{V}$.  The top (bottom) curve is produced using a positive (negative) bias voltage.  The shaded region is the average, and its integral is noted.}
\label{fig:polarity}
\end{figure}

In our testing environment, the fourth of the noted sources of error, the collections of charged particles produced in air surrounding the cup, is our dominant source of uncertainty on total charge in a beam pulse.  Seventy ion-electron pairs are created per electron per cm traveled in air~\cite{leo}.  This means there are plenty of ions or electrons that can hit the core of the cup thereby affecting the measurement.  Each pulse emitted by the accelerator produces roughly the same number of ion-electron pairs.    One partner of these pairs is more likely to strike the Faraday cup with a stronger bias as the bias voltage increases; the polarity dictates whether it is the positive or negative particle.  To first order, we would expect the current due to ionization for positive and negative bias to be equal in magnitude but opposite in sign.  Figure~\ref{fig:polarity} illustrates this by showing the current for different pairs of bias voltages with the same magnitude along with their average, which remains relatively stable under variations of the magnitude of the bias voltage.  Unfortunately, data with varying bias voltage polarity, similar to Figure~\ref{fig:polarity}, are not available for the linac configuration used to irradiate the blocks.  Therefore we cannot use this method to determine the total charge per pulse. 

Instead, we choose to integrate the current as a function of time in Region III of Figure~\ref{fig:6pulses} for the measurement taken with zero bias voltage, where ionization effects are minimal.  When integrating Region III there are two reasonable choices for defining zero: zero voltage drop across the resistor (the dotted line in Figure~\ref{fig:6pulses}) and the current level at the end of Region II, just after the RF noise.  We use the average of these results as our nominal value and assign a systematic error that covers both extremes.

Based on \textsc{Geant4} simulations with zero bias, we estimate that 10\% of the electrons escape due to penetration through the cup and backscatter from the top surface.  We correct for this effect and, conservatively, take the magnitude of this correction as a systematic error.  The measured charge per pulse is -$0.27\pm 0.04~\text{nC}$ where the error is systematic and comes from the integration and leakage correction.  To mitigate pulse-to-pulse variations and electronic noise, the characterization of the pulses was done by averaging the digitized Faraday cup signal for sixty-four consecutive linac pulses.  Irradiation of the blocks was conducted immediately after pulse characterization; effects due to long-time drift of the charge per pulse are assumed to be negligible.
 
\section{Quantifying the Damage}

\begin{figure}
\begin{center}
\includegraphics[width=\linewidth]{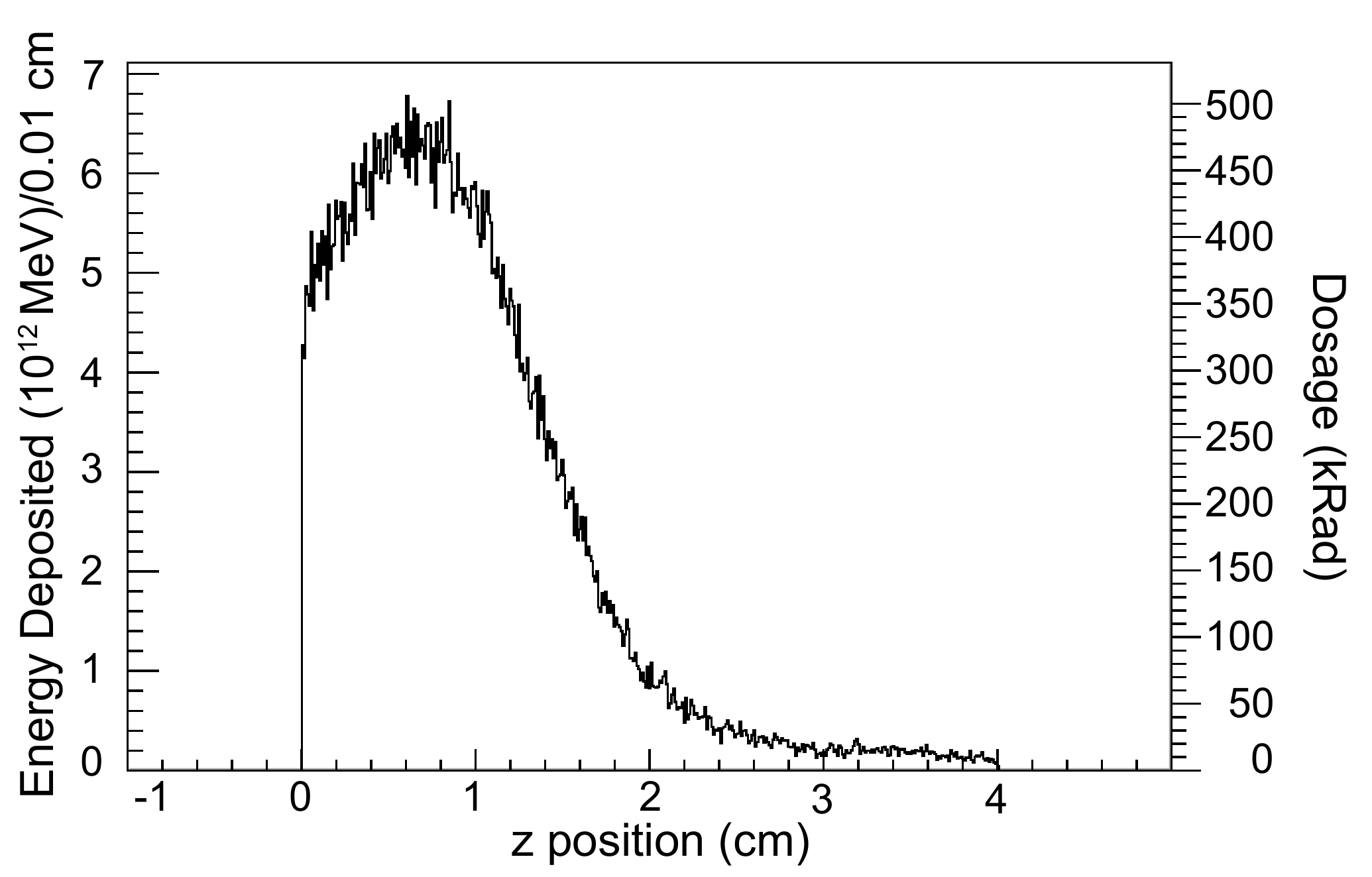}
\caption{\textsc{Geant4} simulation of energy deposition of $10^{14}$ 20 MeV electrons as a function of depth in a $2.4~\cm\times2.5~\cm\times4~\cm$ volume of lead glass.  The dosage scale assumes a volume element of $2.4~\cm\times2.5~\cm\times0.01~\cm$.}
\label{fig:electron}
\end{center}
\end{figure}

The blocks were irradiated with a total (8 to 150)$\times 10^{12}$ electrons with the beam axis perpendicular to the $45~\cm$ dimension of the block.  As noted above the electrons were administered in -0.27~nC pulses at a rate of about 33 pulses per second.  The time to administer the maximum dose was 45 minutes.  Tests with film indicate that the intensity of the beam is uniform across the $2.5~\cm \times 2.4~\cm$ aperture.  Figure~\ref{fig:electron} shows a \textsc{Geant4} simulation of energy deposition per unit volume and mass (kRad) as a function of depth.  The density of F8 lead glass is $3.61~\text{g}/\text{cm}^{3}$.

\begin{figure}
\begin{center}
\includegraphics[width=\linewidth]{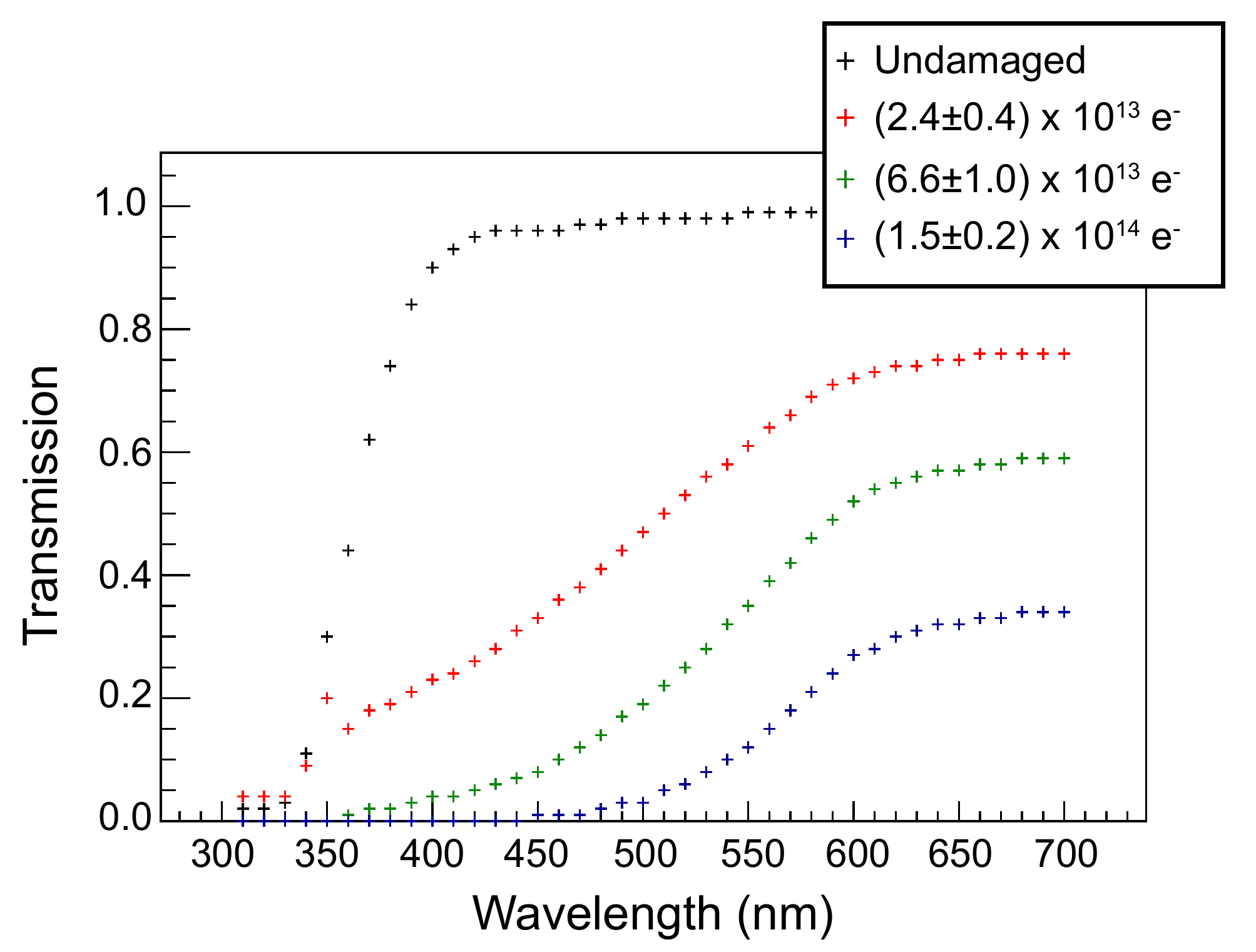}
\caption{Transmission coefficient of $4~\cm$ of lead glass as a function of wavelength for various amounts of radiation.  Estimated errors are 2\% (10\%) for wavelengths above (below) 380 nm.}
\label{fig:tvslam}
\end{center}
\end{figure}

A Shimadzu UV160U spectrophotometer was used to measure the transmission coefficient as a function of wavelength for the irradiated glass.  By examining variations in the results we assign an error of 2\% (10\%) for wavelengths above (below) $380~\nm$.  The data, which have been corrected for reflections at the surfaces, appear in Figure~\ref{fig:tvslam}.  

We seek to characterize the data in Figure~\ref{fig:tvslam} in terms of fundamental properties of lead glass.  We begin with the assumption that the attenuation length of lead glass can be written as a function of the energy per unit volume deposited in the glass as
\begin{equation}
\lambda(\mathcal{E}) = \left[\frac{1}{\lambda_0} + b \mathcal{E}\right]^{-1},
\label{eq:atten}
\end{equation}
where $\lambda_0$ is the attenuation length of undamaged lead glass and $b$ is property of the glass related to how opaque it becomes when irradiated.  Both $\lambda_0$ and $b$ will have a wavelength dependence.

The electron beam flux provided by the linac is uniform in the dimension transverse to the beam axis, and the energy deposited in the glass varies with depth (see Fig.~\ref{fig:electron}).  To analyze the damage, we divide the irradiated region in small volumes with cross sectional area $A$, which matches the beam spot size, and some small depth $z_i$.  The transmission coefficient through such a volume is given by
\begin{equation}
T_i = \exp\left[-\frac{z_i}{\lambda(\mathcal{E}_i)}\right] = \exp\left[-z_i\left( \frac{1}{\lambda_0} + b \mathcal{E}_i \right)\right].
\end{equation}
The transmission coefficient through the entire piece of irradiated glass with thickness or depth $z$ is then given by
\begin{equation}
T = \prod_i T_i = \exp\left[-\frac{1}{\lambda_0}\sum_i z_i\right]\exp\left[-b\sum_iz_i\mathcal{E}_i\right].
\end{equation}
If we note that $\mathcal{E}_i = E_i / (Az_i)$, where $E_i$ is the energy deposited in volume $i$ by the beam, then the transmission coefficient can be rewritten as
\begin{equation}
T(\mathcal{E}) = \exp\left[-\frac{z}{\lambda_0}\right]\exp\left[-bz\mathcal{E}\right].
\label{eq:fiteq}
\end{equation}
Therefore, based on the assumption in Eq.~\ref{eq:atten}, the transmission coefficient through some thickness of glass $z$ should exponentially decrease with increasing energy deposition.  A key feature of Eq.~\ref{eq:fiteq} is that it depends only on the average energy per unit volume deposited in the glass $\mathcal{E}$, assuming it is uniformly deposited in the across an area $A$.

\begin{figure}
\begin{center}
\includegraphics[width=\linewidth]{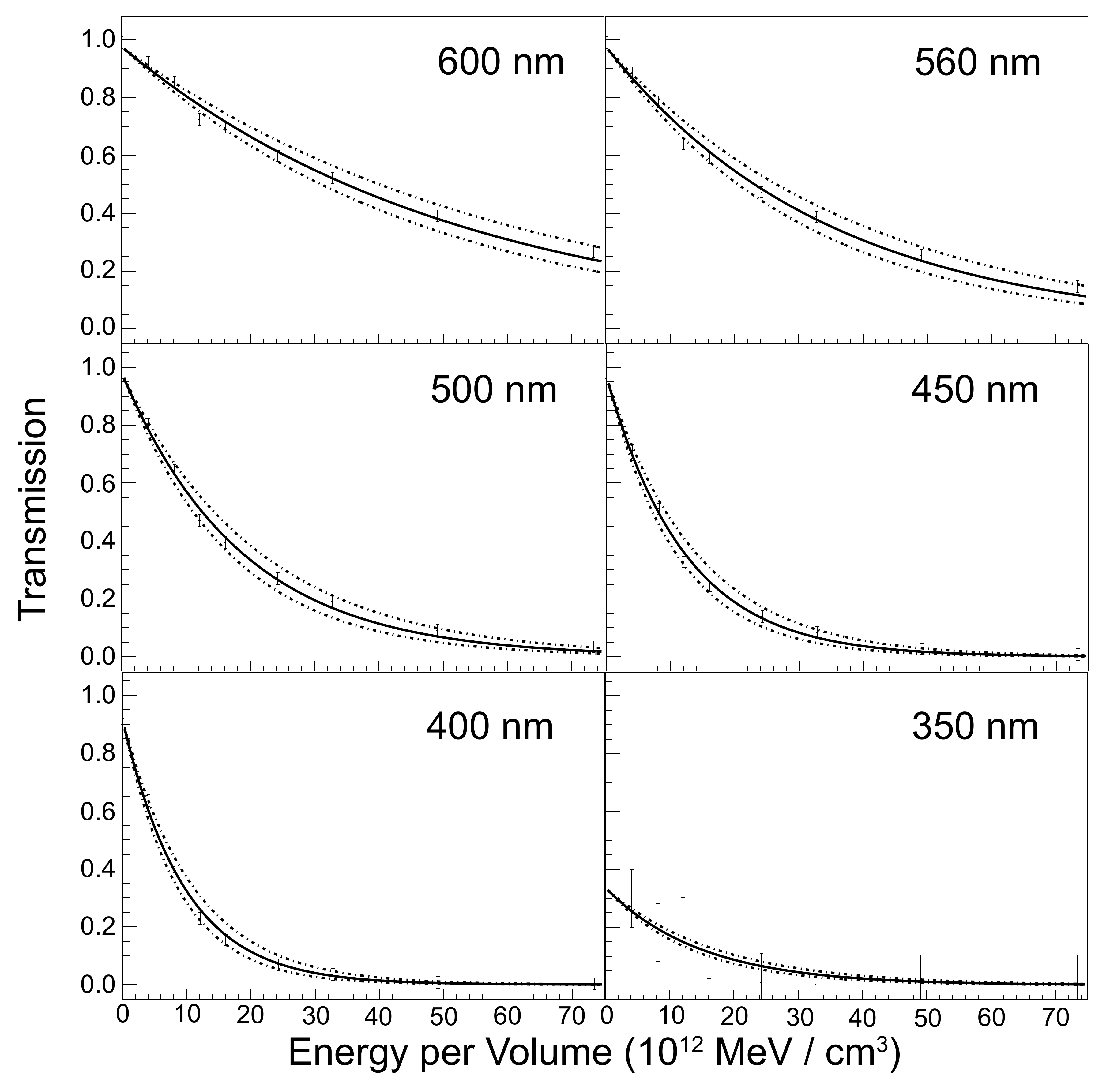}

\caption{Transmission coefficient through $4~\cm$ vs.\ energy per volume ($\mathcal{E}$), fit to Eq.~\ref{eq:fiteq} for different wavelengths.  The solid black line is the  fit to the data using Eq.~\ref{eq:fiteq}.  The dashed lines show the systematic error due to the accelerator calibration, which is correlated among all data points.}
\label{fig:tvd1}
\end{center}
\end{figure}

The measured values of $T$ as a function $\mathcal{E}$ were fit to Eq.~\ref{eq:fiteq} to obtain the parameters $\lambda_0$ and $b$.  The fits are shown in Fig.~\ref{fig:tvd1}, and the fit parameters are compiled in Table~\ref{tab:values}.  The systematic error is due to the uncertainty in the charge per beam pulse delivered by the accelerator.  This error is quantified by reevaluating the data points assuming the minimum and maximum values of the charge per pulse.  We observe that the value of $b$ is smaller for higher wavelengths.  This is consistent with the brownish color of radiation damaged lead glass.
\begin{center}

\begin{table}
\begin{center}
\caption{Results from the fit of data to Eq.~\ref{eq:fiteq}.  The first error is due to variations in spectrophotometer output.  The second error on $b$ is from the uncertainty in the accelerator calibration.\vspace{8pt}}
\label{tab:values}
\begin{tabular}{c|c|c}
\hline\hline
Wavelength& $\lambda_0 (~\cm)$  & $b$ ($~\cm^{2}~/~10^{12} ~\mev$)\\
\hline
		600 nm&	$158_{-54}^{+165} $&	$0.0048\pm0.0002\pm0.0006$\\
		560 nm&	$166_{-62}^{+240}$&	$0.0073\pm0.0003\pm0.0009$\\
		500 nm&	$206_{-96}^{+1186}$&	$0.0134\pm0.0004\pm0.0017$\\
		450 nm&	$135_{-51}^{+210}$&	$0.0205\pm0.0007\pm0.0027$\\
		400 nm&	$47.8_{-9.2}^{+14.7}$&	$0.0261\pm0.0002\pm0.0034$\\
		350 nm&	$3.68_{-0.73}^{+0.94}$&	$0.0168\pm0.0074\pm0.0022$\\

\hline \hline

\end{tabular}

\end{center}
\end{table}
\end{center}

\section{Comparison to Proton Beam}

\begin{figure}
\begin{center}
\includegraphics[scale=.4]{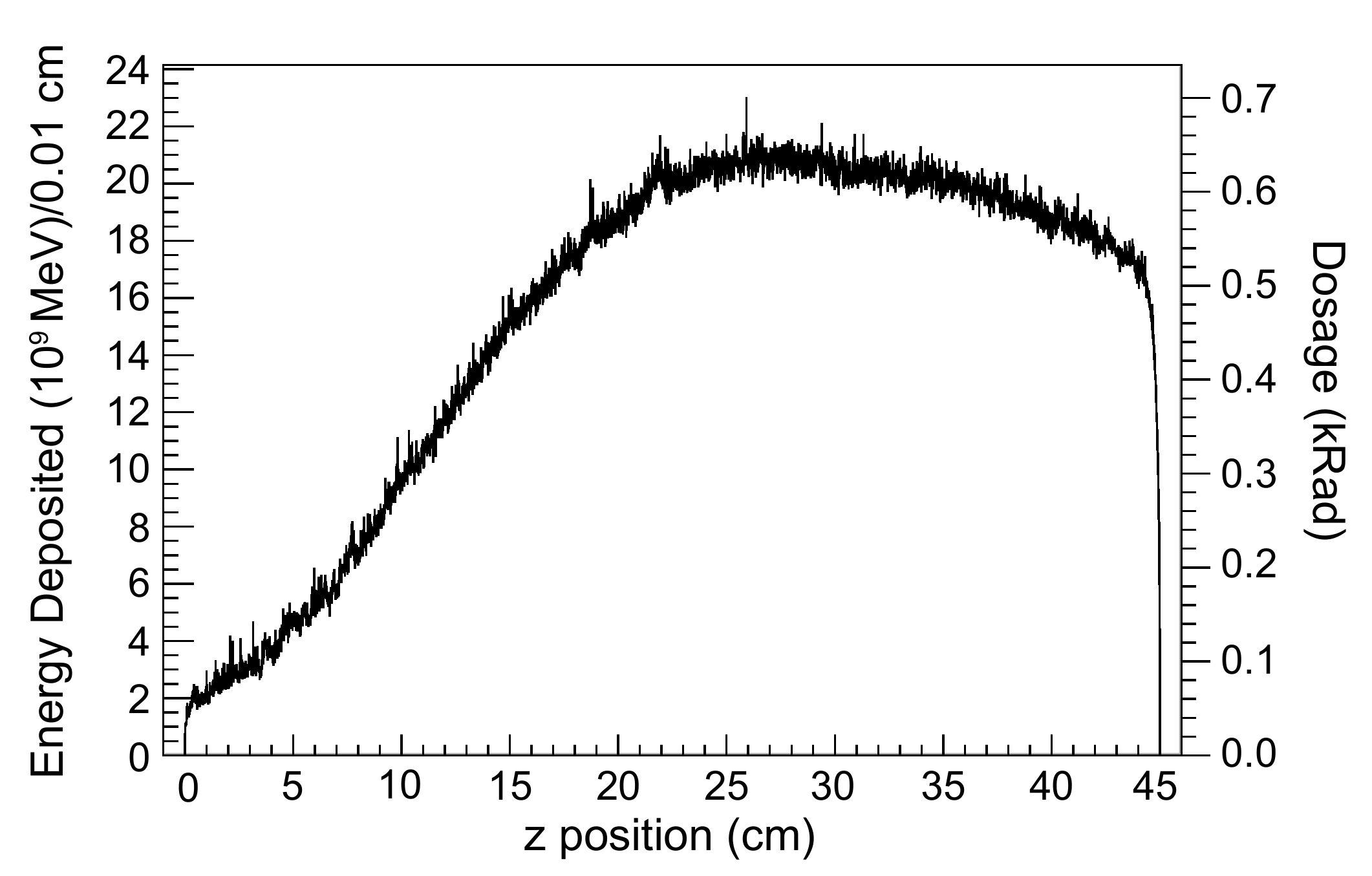}
\caption{\textsc{Geant4} simulation of energy deposition of $10^{10}$ protons with energy of $70~\gev$ incident on $3.8~\cm\times3.8~\cm\times45~\cm$ F8 lead glass using the quark-gluon string precompound model (QGS/Preco)~\cite{qgs}.  The dosage scale assumes a volume element of $3.8~\cm\times3.8~\cm\times0.01~\cm$.}
\label{fig:proton}
\end{center}
\end{figure}

A.V. Inyakin~{\it et al.}\ have studied F8 lead glass that was irradiated with $70~\gev$ protons~\cite{damage}.  The transmission of light through the block was measured before and after irradiation using green and yellow colored LEDs.  A value of $b$ for their data can be obtained by accounting for the average energy deposited per proton using a \textsc{Geant4} simulation (see Fig.~\ref{fig:proton}).  The results are $b = (0.062\pm0.012) \times ($cm$^{2}/10^{12} ~\mev)$ for the green LED  and $b = (0.047\pm0.009) \times($cm$^{2}/10^{12} ~\mev) $ for the yellow LED.  Comparing these results to those for $560~\nm$ and $600~\nm$ respectively, we note that the values of $b$ are considerably larger than our results obtained with an electron beam.  We conclude that F8 lead glass is more susceptible to proton damage than electron damage.  Several studies on different materials support our finding by noting a difference between photon-induced and proton-induced damage~\cite{support,support1,support2}.  

\section{Summary}
Motivated by the design of the forward calorimeter for the GlueX experiment, we present a study of radiation damage due to $20~\mev$ electrons in F8 lead glass.  Using a formalism that models the decrease in attenuation length with increased dose, we successfully fit transmission coefficients as a function of the density of deposited energy.  This formalism then provides a quantitative prediction for the attenuation length as a function of dosage for a variety of wavelengths.  Our measurements confirm the expectation based on visual evidence that, when the glass is irradiated, the attenuation length in the blue portion of the spectrum is degraded more rapidly than in the red portion of the spectrum.  A comparison of our results to those obtained utilizing a hadron beam indicates that approximately equivalent damage to the glass occurs when the total energy deposited by the hadron beam is one tenth of that deposited by the electromagnetic beam.

\section{Acknowledgments}

We would like to thank the staff of the IUCEEM and the Advanced eLectron Photon fAcility (ALPHA), particularly P.~Sokol, for allowing us to utilize the Varian Clinac to irradiate the glass samples.  We would also like to acknowledge helpful discussions with P.~Cole.  This work was supported by the US Department of Energy Office of Nuclear Physics under award DE-FG02-05ER41374.

\end{document}